\documentclass[a4paper,11pt]{article}
\usepackage{pos}
\usepackage{amsmath}
\usepackage{xspace}
\usepackage[capitalise]{cleveref}

\crefname{figure}{Fig.}{Figs.}
\Crefname{figure}{Figure}{Figures}

\title{Differential probes of jet evolution with charged energy-energy correlators at ALICE}

\author*[a]{Tucker Hwang}
\onbehalf{on behalf of the ALICE Collaboration}
\affiliation[a]{University of California, Berkeley\\
  Berkeley, CA, USA}

\emailAdd{tucker\_hwang@berkeley.edu}

\newcommand{\RL}{R_\mathrm{L}}

\newcommand{\pt}{p_\mathrm{T}}

\newcommand{\ptjet}{p_\mathrm{T, ch\ jet}}
\newcommand{\pPb}{\mbox{p--Pb}\xspace}
\newcommand{\dif}{\ensuremath{\mathrm{d}}\xspace}
\newcommand{\sigeec}{\ensuremath{\Sigma_\mathrm{EEC}}\xspace}

\abstract{Energy-energy correlators (EECs) are energy-weighted cross-sections of particle pairs. They reflect QCD dynamics across the full scales of jet evolution by separating energy scales in the jet fragmentation through the angular distance of the particle pairs in the jet. Charged EECs probe the energy flux carried by pairs with the same or opposite electric charges. The interplay between energy distribution and charge conservation enables charged EECs to provide novel constraints on hadronization mechanisms. We present measurements of two-point charged EECs for charged-particle jets in pp collisions at $\sqrt{s} = 5.02$ TeV using the ALICE detector, and compare them to models employing different parton shower and hadronization prescriptions. We also present the first measurement of charged EECs in \pPb collisions, examining cold nuclear matter effects on jet dynamics.}

\FullConference{13th International Conference on Hard and Electromagnetic Probes of High-Energy Nuclear Collisions (HardProbes2026)\\
21-26 June 2026\\
Nashville, Tennessee, USA\\}

\begin{document}
\maketitle

\section{Introduction}
\label{intro}
Originally an event shape observable for precision measurements of the strong coupling constant from early $e^+e^-$ experiments \cite{alphas}, the energy--energy correlator has seen a recent resurgence at hadron colliders. It is defined as an energy-weighted two-particle correlation function:
\begin{equation}
\label{eq:eec}
    \sigeec(\RL) = \frac{1}{ N_{\mathrm{jet}}\cdot\Delta}\int_{\RL-\frac{1}{2}\Delta}^{\RL+\frac{1}{2}\Delta}\dif\RL' \,\, \sum_{\mathrm{jets}} \,\,\, \sum_{i, j \in \mathrm{jet}} \frac{p_{\mathrm{T},i}}{\ptjet}\frac{p_{\mathrm{T},j}}{\ptjet} \delta(\RL'-R_{\mathrm{L}, ij}),
\end{equation}
where $i$ and $j$ run over jet constituents, $N_\mathrm{jet}$ is the number of jets, $\Delta$ is the angular bin width, $\ptjet$ is the jet transverse momentum, $R_{\mathrm{L},ij}$ is the angular distance between particles $i$ and $j$, and $\delta$ is the Dirac delta function. Due to the angular ordering of a jet's parton shower, the EEC probes an energy scale (or virtuality) $Q$ of the parton splittings proportional to $Q \sim \ptjet\RL$ \cite{eec}. Thus, at large angles, the EEC probes splittings early in the parton shower. At small angles, the jet constituents do not interact; their energies and angular separation are uncorrelated, and thus they are uniformly distributed in angle. In between these two regimes, it is believed that the EEC encodes some information about the hadronization transition. The EEC has been measured in a variety of contexts, from the Relativistic Heavy Ion Collider~\cite{star} to the Large Hadron Collider; from small to small~\cite{star,aliceeecpp,cmspp,aliceppb,atlasteec} to large collision systems~\cite{cmspbpb}, and in inclusive and D$^0$ meson-tagged jets~\cite{aliceeecd0}.

\section{The charged energy--energy correlator}

The charged EECs \cite{ceec_th} extend \Cref{eq:eec} to also track the flow of charge as well as energy. The \textit{charge-selected correlators}, $\Sigma_\mathrm{EEC}^{+-}$ (the unlike-sign EEC), $\Sigma_\mathrm{EEC}^{++}$ and $\Sigma_\mathrm{EEC}^{--}$ (the like-sign EECs) are defined similarly to \Cref{eq:eec}, but only pairs $(i, j)$ of unlike-sign and like-sign charges are considered, respectively. The \textit{charge-weighted correlator} $\Sigma_\mathrm{EEC}^{\mathcal{Q}}$ is defined as
\begin{equation}
\label{eq:ceec-wt}
    \sigeec(\RL) = \frac{1}{ N_{\mathrm{jet}}\cdot\Delta}\int_{\RL-\frac{1}{2}\Delta}^{\RL+\frac{1}{2}\Delta}\dif\RL' \,\, \sum_{\mathrm{jets}} \,\,\, \sum_{i, j \in \mathrm{jet}} q_iq_j\frac{p_{\mathrm{T},i}}{\ptjet}\frac{p_{\mathrm{T},j}}{\ptjet} \delta(\RL'-R_{\mathrm{L}, ij}),
\end{equation}
where the product of the charges are included in the energy weight. Note that the charge-weighted correlator can be rewritten as $\Sigma_\mathrm{EEC}^{\mathcal{Q}} = \Sigma_\mathrm{EEC}^{++} + \Sigma_\mathrm{EEC}^{--} - \Sigma_\mathrm{EEC}^{+-}$. The charge-weighted correlator quantifies the overall balance between like-sign and unlike-sign pairs, with like-sign pairs contributing positive weights and vice versa.\footnote{This formalism breaks when considering doubly charged particles, but their contribution to jets is negligible.}

By investigating energy flow differentially in charge and angle, measurements of charged EECs provide strong constraints on hadronization models. In addition, EEC measurements in proton--lead (p--Pb) collisions have shown modifications relative to the proton--proton (pp) baseline \cite{aliceppb}. Charged EECs are a natural extension of the EEC that can probe the underlying source of these differences, and serve as a new baseline for models.

\section{Experimental setup and analysis techniques}

This measurement \cite{aliceceec} mainly utilizes three subdetectors of the ALICE experiment: the Inner Tracking System (ITS) and the Time Projection Chamber (TPC) track charged particles with $|\eta| < 0.9$, and the V0 scintillator hodoscopes provide the minimum-bias trigger \cite{perf1,perf2}. ALICE is well equipped for studies of jets at low $\pt$ due to its high tracking efficiency and good momentum resolution at low momenta.

The data were collected in Run 2 at $\sqrt{s_\mathrm{NN}}=5.02$ TeV. Anti-$k_\mathrm{T}$ jets \cite{antikt} with $R=0.4$ are constructed via $E$-scheme recombination from charged particles with $\pt>150$ MeV/$c$. Only jets with their full area within the TPC fiducial acceptance (i.e. $|\eta_\mathrm{jet}| < 0.9 - R = 0.5$) are considered. In pp collisions, EEC pairs are only constructed from tracks with $\pt > 1$ GeV/$c$ to reduce the contribution from the underlying event (UE). In p--Pb collisions (and any corresponding pp reference comparisons), a jet area-based $\ptjet$ correction \cite{rho} and a perpendicular cone subtraction (that removes contributions from pairs that include a particle from the UE) are also included \cite{aliceppb,aliceceec}.

The data are corrected for detector effects with \textsc{Pythia 8} (DPMJET) simulations of pp (p--Pb) collisions, propagated through a \textsc{Geant3} representation of the ALICE detector. Because the single-track angular resolution is much smaller than the $\RL$ bin widths, no angular bin migrations are considered in the corrections \cite{aliceeecpp}. Thus, a bin-by-bin correction is employed to account for corrections to $\ptjet$ and energy weights.

Systematic uncertainties arise from the correction for detector effects and, in the case of the \pPb results, the subtraction of contributions from the underlying event. Uncertainties from the correction include uncertainties on the reconstruction efficiency of single tracks and track pairs, the dependence of the correction on the event generator, and the correction procedure itself (i.e. a comparison to a more traditional Bayesian unfolding technique).

\section{Results and discussion}
\begin{figure}[!htb]
\centering
\includegraphics[width=0.95\textwidth]{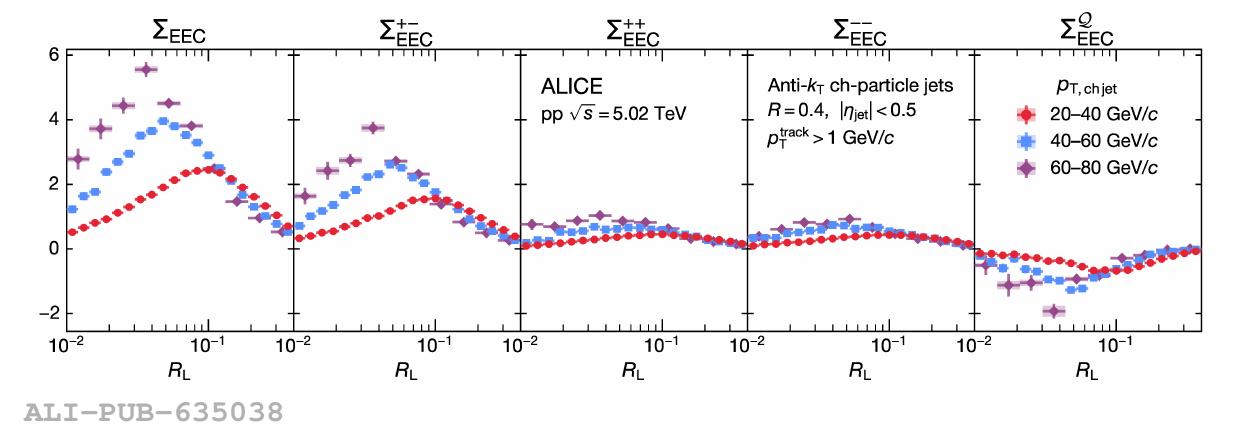}
\caption{The charge-inclusive and charged EECs in pp collisions at $\sqrt{s} = 5.02$ TeV.}
\label{fig:eec-pp}
\end{figure}%
The charged and inclusive EECs in pp collisions are shown in \cref{fig:eec-pp}. In general, the charged EECs show similar shapes to the inclusive EEC. A larger contribution from unlike-sign pairs to the charge-inclusive EEC than like-sign pairs is observed via the negativity of the charge-weighted EEC.

\begin{figure}[!htb]
\centering
\includegraphics[width=0.95\textwidth]{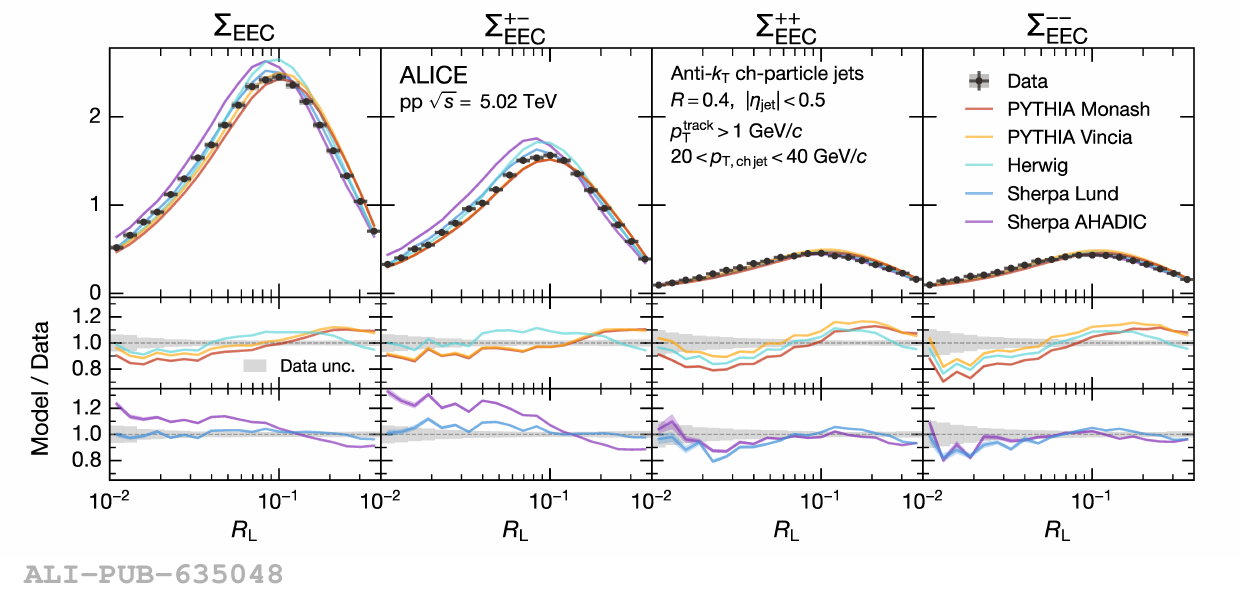}
\caption{Comparison of charged EECs in pp with five different Monte Carlo models.}
\label{fig:eec-mc}
\end{figure}%

The comparisons in \cref{fig:eec-mc} test how jet evolution is modeled in MC generators. Because unlike-sign pairs dominate the inclusive EEC, differences between models in the inclusive EEC follow those in the unlike-sign, similar to what is observed in \textsc{Pythia} and \textsc{Herwig}. Two identical string hadronization models but with different parton shower implementations, \textsc{Pythia} Monash and Vincia, show variations only in like-sign correlations, whereas the opposite is true for identical shower modeling and independent hadronization models (Sherpa Lund and AHADIC). No models coherently reproduce all the data, with no particular model being favored over the others.

\begin{figure}[!htb]
\centering
\includegraphics[width=0.95\textwidth]{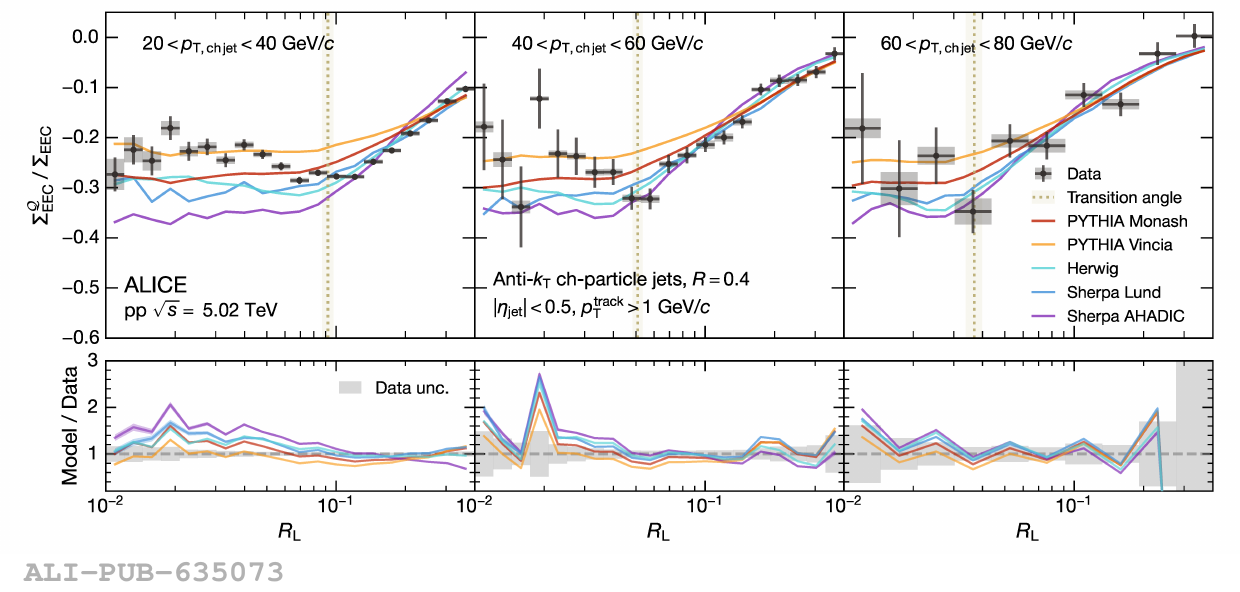}
\caption{The ratio of the charge-weighted to inclusive EEC in pp collisions.}
\label{fig:asym}
\end{figure}%

By normalizing the charge-weighted EEC by the inclusive EEC, the resulting \textit{charged EEC asymmetry} directly (\cref{fig:asym}) quantifies the relative contributions of unlike-sign and like-sign pairs to the inclusive EEC. For example, a value of $-0.4$ means that unlike-sign and like-sign pairs contribute 70\% and 30\%, respectively, to the inclusive EEC. Comparisons of the ALICE data to MC models show an increasing asymmetry (farther from zero) as the angle decreases. In the data, this asymmetry then turns back slightly closer to zero past the transition angle (the angle corresponding to the maximum of the charged EECs in \cref{fig:eec-pp}) in 20--40 GeV/$c$ jets. Despite the pronounced differences among the MC models in \cref{fig:eec-mc}, they all predict a flat asymmetry past this angle. The MC generators predict the same behavior at higher jet $\pt$, but it is not possible confirm this in the data due to insufficient statistical precision.

\begin{figure}[!htb]
  \begin{minipage}[c]{0.6\textwidth}
    \includegraphics[width=0.9\textwidth]{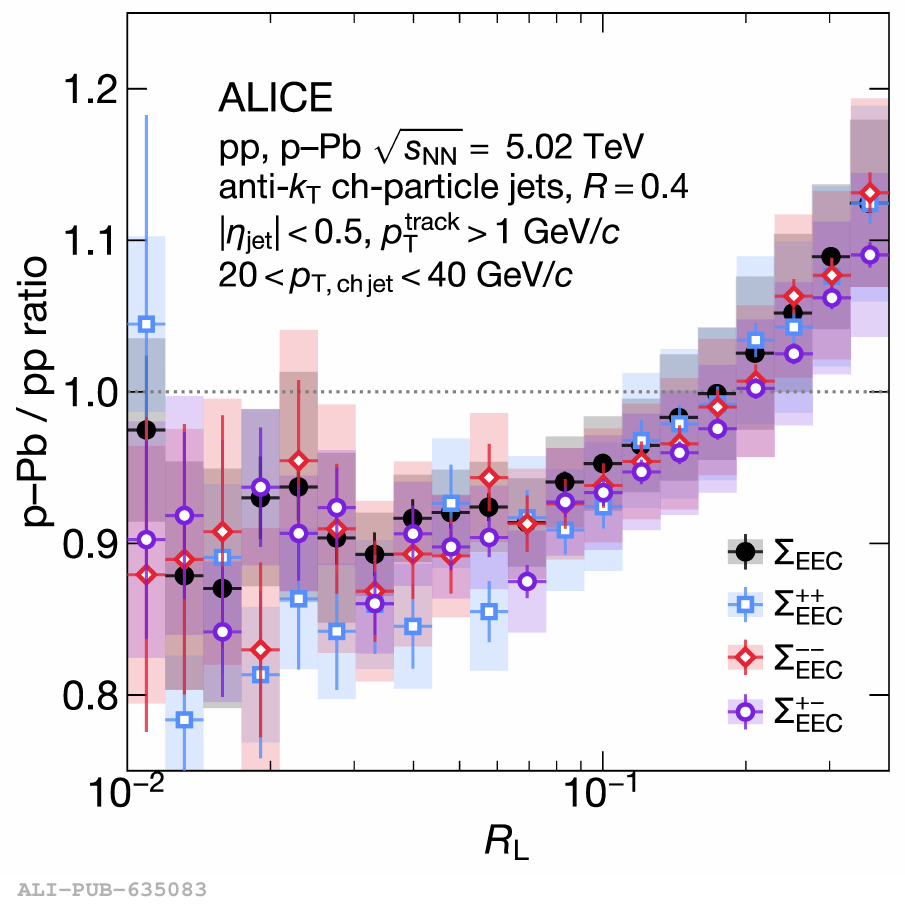}
  \end{minipage}\hfill
  \begin{minipage}[c]{0.4\textwidth}
    \caption{The ratio of charged EECs in p--Pb to pp collisions for inclusive jets with $20 < \ptjet < 40$ GeV/$c$.}
    \label{fig:eec-ratio}
  \end{minipage}
\end{figure}

ALICE has recently measured the inclusive EEC in p--Pb collisions and noted an enhancement at large angles relative to pp collisions \cite{aliceppb}. \Cref{fig:eec-ratio} shows that this enhancement is not dependent on charge.

\section{Conclusions and future prospects}

Charged EECs can track energy and charge flow in jets, providing increased sensitivity to hadronization and shower mechanisms compared to the inclusive EEC. No MC generators are able to fully describe the data; in particular, they fail to qualitatively describe the behavior of the charged EEC asymmetry at small angles. Extending into p--Pb collisions, the charge-selected EECs have the same modification in cold nuclear matter as the inclusive EEC. The heightened sensitivity of charged EECs motivates further studies, in large collision systems (e.g. Pb--Pb) and a wider range of jet samples, such as a quark-enriched jet sample identified via jets recoiling from a photon.

%
\bibliographystyle{JHEP}
\bibliography{refs.bib} 
%
%
%
%



\end{document}